\documentclass[%
reprint,
superscriptaddress,
amsmath,amssymb,
aps,
prl,
floatfix
]{revtex4-2}

\usepackage{multirow}
\usepackage{graphicx}
\usepackage{dcolumn}
\usepackage{bm}
\usepackage[version=3]{mhchem}
\usepackage{nicefrac}

\usepackage{color}
\usepackage{xcolor}
\usepackage{dcolumn}
\usepackage{amssymb}
\usepackage{url}
\usepackage{hyperref}
\usepackage{xfrac}
\usepackage{tabularx}
\usepackage{xspace}
\usepackage{amsmath}
\usepackage[normalem]{ulem}
\usepackage{mathtools}
\usepackage{physics}
\usepackage{dsfont}
\usepackage{float}
\usepackage{braket}
\usepackage{placeins}
\usepackage{footmisc}
\usepackage{multirow}
\usepackage{textcomp}

\def\be{\begin{equation}}
\def\ee{\end{equation}}
\def\bea{\begin{eqnarray}}
\def\eea{\end{eqnarray}}
\def\ba{\begin{array}}
\def\ea{\end{array}}

\begin{document}
\title{Theory of the $1/9$ Magnetization Plateau of the Spin-$1/2$ Kagome Heisenberg Antiferromagnet}
\author{Pratyay Ghosh}
\email{pratyay.ghosh@epfl.ch}
\affiliation{Institute of Physics, Ecole Polytechnique Fédérale de Lausanne (EPFL), CH-1015 Lausanne, Switzerland}

\author{Fr\'ed\'eric Mila}
\affiliation{Institute of Physics, Ecole Polytechnique Fédérale de Lausanne (EPFL), CH-1015 Lausanne, Switzerland}

\begin{abstract}
We investigate the $1/9$ magnetization plateau of the spin-$1/2$ kagome Heisenberg antiferromagnet using infinite projected entangled-pair states and identify a spin-density-wave state with a $27$-site unit cell and a lower variational energy than previous approaches. In contrast to higher-magnetization plateaus, which are well described by localized magnons on nonoverlapping hexagons and preserve a sixfold rotational symmetry, the $1/9$ plateau only keeps a threefold rotational symmmetry and consists of a tiling by three types of $9$-site clusters related by $\pm2\pi/3$ rotations. These conclusions are supported by a complementary cluster mean-field theory which correctly preserves the symmetry of the higher plateaus but leads to the same spontaneous symmetry breaking for the $1/9$ plateau with a qualitatively similar pattern of nine different local magnetizations.
\end{abstract}

\maketitle
The spin-$1/2$ kagome Heisenberg antiferromagnet (KHA) is a paradigmatic example of a frustrated quantum magnet, in which strong frustration and quantum fluctuations suppress conventional magnetic order and produce a dense manifold of low-energy states~\cite{Diepbook,frustrationbook,PhysRevB.47.5459,kagome_mila}, making it one of the most intensively studied systems in quantum magnetism. The nature of its ground state remains under debate, with leading theoretical proposals including quantum spin liquids (QSLs) with fractionalized excitations and valence-bond crystals (VBCs)~\cite{PhysRevB.68.214415,PhysRevLett.98.117205,PhysRevB.76.180407,Evenbly2010,PhysRevLett.101.117203,PhysRevLett.109.067201,Yan2011,PhysRevLett.118.137202,PhysRevX.15.011047}.The application of a magnetic field further enriches the phase diagram, as the magnetization exhibits a sequence of fractional plateaus that signal distinct field-induced quantum phases~\cite{PhysRevLett.88.167207,Okamoto2011,Capponi2013,Nishimoto2013,Ishikawa2015,Picot2016,Yoshida2017,Chen2018,Okuma2019,Yoshida2022,Fang2023,He2024,Morita2024,Kato2024,Jeon2024,Suetsugu2024,He2025,Cheng2026,Fang2026,Raikos2026,Duric2026}. The higher-field plateaus at $m/m_{s}=1/3$, $5/9$, and $7/9$, where $m_{s}$ denotes the saturation magnetization, are relatively well understood. In contrast, the microscopic nature of the low-field $1/9$ plateau remains unresolved and is the subject of ongoing debate.

High-field measurements conducted in 2024 on yttrium-based hydroxyhalide kagome antiferromagnets confirmed the presence of a $1/9$ magnetization plateau~\cite{Jeon2024,Suetsugu2024,q3yj-t2hj,Zheng2025,hayashi202659magnetizationplateauspinsupersolidity}.
A $1/9$ magnetization plateau has also been reported recently in \ce{Y}-kapellasite~\cite{chatterjee2026gapped19magnetizationplateau}.
Further experimental evidence for plateau-like behavior has emerged from high-field measurements of the titanium-based kagome family $\ce{Cs8AB3Ti12F48}$ (\ce{A} = \ce{Rb}, \ce{Li}; \ce{B} = \ce{K}, \ce{Na})~\cite{chaudhary2026universality19magnetizationplateau};
both $\ce{Cs8RbK3Ti12F48}$ and $\ce{Cs8LiK3Ti12F48}$ exhibit robust $1/9$ plateau-like phases, despite the absence of the typically more prominent $1/3$ plateau. In contrast, $\ce{Cs8LiNa3Ti12F48}$ displays suppression of both the $1/9$ feature and the quantum-disordered zero-field ground state. These findings indicate that the $1/9$ plateau is sensitive to the underlying frustrated exchange network. In the ideal KHA, the higher-field $1/3$, $5/9$, and $7/9$ plateaus are attributed to localized magnons~\cite{Richter2004_magnon,Richter2005}.
Bond anisotropies disrupt this mechanism by introducing dispersion to the localized magnons, which can suppress these plateaus. The persistence of the $1/9$ plateau in titanium-based compounds, despite deviations from the ideal kagome model~\cite{chaudhary2026universality19magnetizationplateau}, suggests the presence of a qualitatively different mechanism that is more robust against bond anisotropies. However, the microscopic state responsible for this plateau remains unresolved. Theoretical studies of the ideal spin-$1/2$ KHA propose several competing states: a topological $\mathbb{Z}_3$ QSL~\cite{Nishimoto2013,He2024}, VBCs with $\sqrt{3}\times\sqrt{3}$~\cite{Picot2016,Fang2023,Fang2026} or $3\times3$ periodicity~\cite{Cheng2026,Raikos2026}, and a spin-density-wave (SDW) state~\cite{Duric2026}.

In this Letter, we investigate the $1/9$ magnetization plateau of the spin-$1/2$ KHA using infinite projected entangled-pair states (iPEPS)~\cite{Jiang2008,Jordan2008,Verstraete2004a,Verstraete2004b,Gu2008} and the corner transfer matrix renormalization group (CTMRG)~\cite{Nishino1996,Orus2009,star_ghosh,Nyckees2026,Ruby_Ghosh}. Extending previous tensor network studies, which were restricted to $9$-site unit cells~\cite{Picot2016,Fang2023,Fang2026}, we consider a $27$-site unit cell. Our results demonstrate that the $1/9$ plateau is stabilized over a finite range of applied magnetic fields. We identify the underlying state as an SDW with a $27$-site unit cell composed of three types of $9$-site clusters related by successive $2\pi/3$ rotations. This structure breaks the sixfold rotational symmetry of all hexagons and only keeps three inequivalent centers of threefold rotation. Both the nearest-neighbor (NN) spin correlations and the local magnetizations display spatial modulation, with nine different local magnetizations. A complementary cluster mean-field theory (CMFT) analysis reveals the same spontaneous breaking of sixfold rotational symmetry and produces a qualitatively consistent magnetization pattern. 

\begin{figure}[t]
    \includegraphics[width=0.95\columnwidth]{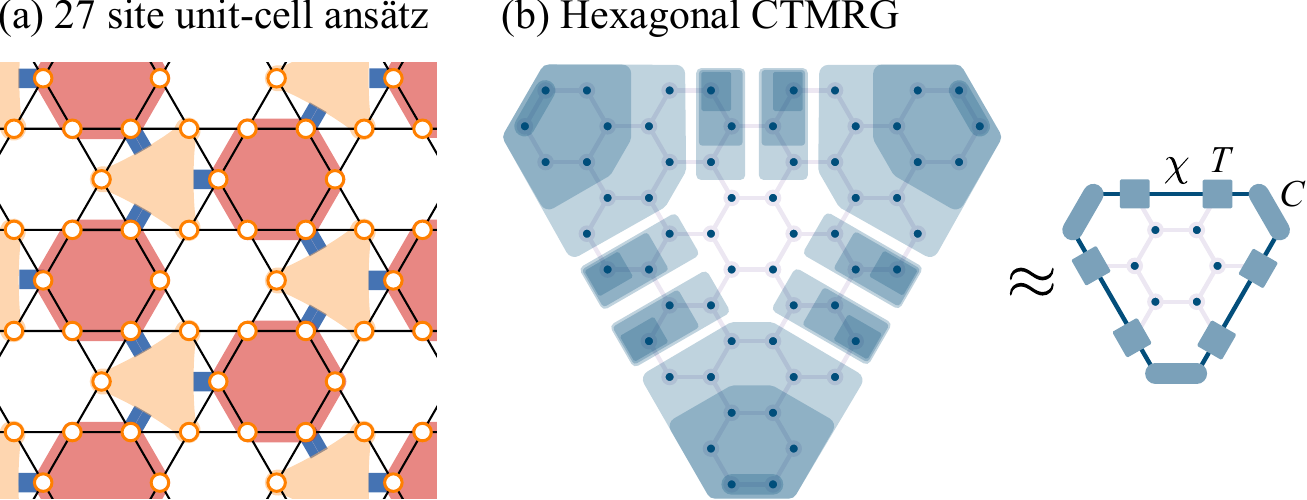}
    \caption{The tensor network with a $27$-site unit cell used in the iPEPS calculations. The network forms a honeycomb geometry, with tensors connected by virtual bonds depicted by thick blue lines. The infinite environment surrounding the unit cells is addressed using the hexagonal CTMRG scheme~\cite{Nyckees2023,star_ghosh,Nyckees2026,Ruby_Ghosh}. (b) The schematic illustrates the construction of the hexagonal CTMRG environment, which consists of three corner tensors $C$ and six edge tensors $T$ surrounding the central region. These tensors have auxiliary bonds of dimension $\chi$.
    }
 \label{fig-TN}
\end{figure}

Previous tensor network studies employed $9$-site unit cells, which represent the minimum size compatible with the $1/9$ magnetization state~\cite{He2024,Picot2016,Fang2023,Fang2026}. The variational energies obtained in these studies are comparable to those of the proposed $\mathbb{Z}_3$ QSL.~\cite{Nishimoto2013}. More recent calculations utilizing resonating-valence-bond (RVB) wave functions~\cite{Cheng2026} and neural-network quantum states (NQS)~\cite{Raikos2026} have achieved lower energies with a $27$-site unit cell. These finite-system results, particularly the size-dependent outcomes observed in NQS, motivate an investigation of the $1/9$ plateau in the thermodynamic limit. The present study examines infinite systems with $27$-site variational unit cells, referred to as the $27$-site ansatz, utilizing the tensor constructions shown in Fig.~\ref{fig-TN}(a). A $36$-site unit cell construction was also considered; however, stabilization of the $1/9$ plateau could not be achieved within this ansatz. The tensor network representing the $27$-site ansatz consists of three six-site and three three-site coarse-grained tensors. The size of the variational unit cell does not determine the periodicity of the state; larger cells can accommodate both smaller-period states and structures with periods up to the chosen size. These constructions enable the examination of spatial structures proposed in earlier studies. In this construction, the tensors form a honeycomb lattice connected by virtual bonds of dimension $D$, allowing for analysis of the infinite two-dimensional environment using the hexagonal CTMRG scheme~\cite{Nyckees2023,star_ghosh,Nyckees2026,Ruby_Ghosh}.

Due to the high computational cost associated with full-update~\cite{Jordan2008,Phien2015} and automatic-differentiation~\cite{Hasik2021} optimizations for enlarged unit cells, we employ simple update~\cite{Jiang2008} to access larger bond dimensions. Since our primary focus is on the magnetization plateau, which is expected to be gapped, the simple-update approximation is expected to capture its essential physics.
The resulting infinite tensor networks are contracted using the hexagonal CTMRG environment, which consists of three corner tensors $C$ and six edge tensors $T$ [see Fig.~\ref{fig-TN}(b)].
The auxiliary dimension $\chi$ determines the accuracy of the approximate contraction.
We verify convergence of energy and magnetization with respect to $\chi$.
Initial optimization with a six-tensor unit cell produces states that approximately preserve threefold rotational symmetry about the center across the entire field range studied. By exploiting this symmetry, a reduced two-tensor representation is employed in subsequent calculations~\cite{star_ghosh,Ruby_Ghosh}, thereby improving numerical convergence while preserving the structure of the optimized state.

\begin{figure}[t]
    \includegraphics[width=0.95\columnwidth]{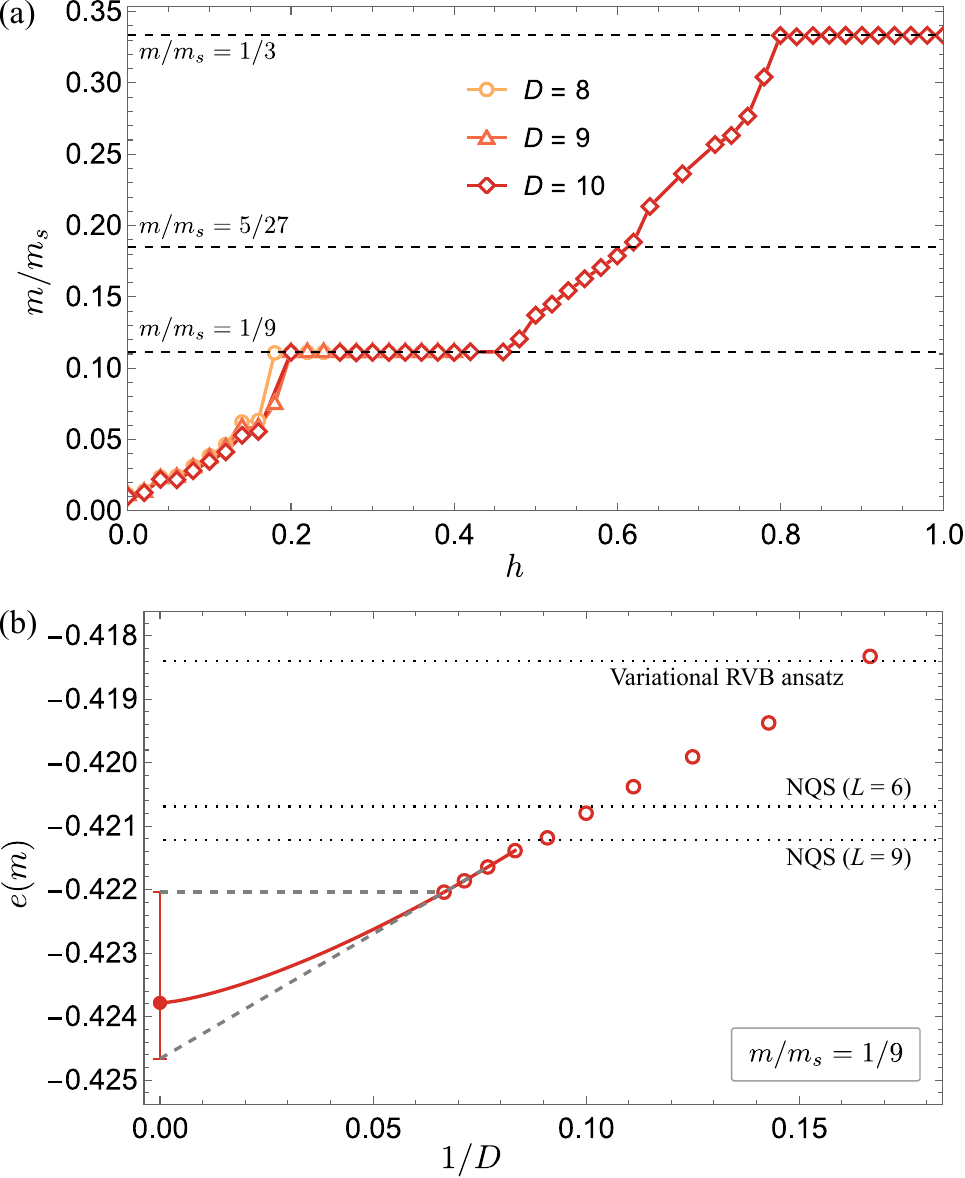}
    \caption{(a) Magnetization, $m/m_{s}$, as a function of $h$ for $D=8,9,10$. Within the 27-site ansatz, the magnetization increases continuously from $h=0$, and the $1/9$ plateau is stabilized for $h\in(0.2,0.46)$. This is followed by a continuous increase toward the $1/3$ plateau, with a weak feature observed near $m/m_{s}=5/27$. (b) Energy per site, $e(m)=e(m,h)-hm$, of the $1/9$ plateau state as a function of bond dimension $D$. Solid lines represent large-$D$ algebraic extrapolations, and conservative error bars denote the estimated uncertainty based on the largest-$D$ energy and a linear extrapolation of the last three data points. Horizontal lines indicate previous RVB~\cite{Cheng2026} and NQS~\cite{Raikos2026} results. The $D\ge12$ iPEPS energy is lower than all previously reported values. The previously reported $9$-site iPEPS~\cite{Fang2023} and $Z_3$ QSL~\cite{He2024} energies are outside the displayed range. 
} \label{fig-energy}
\end{figure}

In the presence of a magnetic field, the Hamiltonian includes a Zeeman term, $-h\sum_i S_i^z$. The field $h$ is expressed in units of the nearest-neighbor (NN) exchange. Figure~\ref{fig-energy}(a) presents magnetization curves derived from the $27$-site ansatz for various bond dimensions $D$ as a function of $h$. The magnetization increases continuously from zero at zero field and attains the $1/9$ plateau at $h\simeq0.2$. This plateau persists up to $h\simeq0.46$, after which the magnetization rises toward the $1/3$ plateau, with a subtle feature near $m/m_s=5/27$. Figure~\ref{fig-energy}(b) displays the energy $e(m)=e(m,h)-hm$ of the $m/m_s=1/9$ state as a function of $D$. At $D\geq12$, the calculated energy is already lower than those obtained from both RVB wave-function ansatz~\cite{Cheng2026} and NQS~\cite{Raikos2026}. Extrapolation to large $D$ yields an even lower value [see also Tab.~\ref{tab:comparison}]. In contrast, previously reported energies for $9$-site iPEPS and iPESS~\cite{Fang2023}, as well as for the proposed $\mathbb{Z}_3$ QSL state~\cite{He2024}, are substantially higher~\footnote{We exclude Ref.~\cite{Duric2026} from the quantitative comparison because its reported estimate of $e(m)$ for the $1/9$ plateau is lower than established estimates of the zero-field KHA ground-state energy, including the value reported in their own previous work~\cite{PhysRevX.15.011047}. This discrepancy also appears to conflict with an exact lower bound on the zero-field energy~\cite{Raikos2026}, and its origin remains unclear.}. Although the energy is not fully converged at the largest $D$ considered, its consistent decrease with increasing $D$, along with the lower energy already achieved at $D\ge12$, provides strong evidence supporting the state obtained with the $27$-site ansatz.

\begin{table}[]
\centering
\begin{tabular}{l|l|l|l|l|}
Candidate state                                 & Method                          &                   & $e(m)$                   & Field range                             \\ \hline
$\mathbf{Z}_3$ QSL                              & VMC                             &                   & -0.41178                 & $[0.35,0.63]$                           \\ \hline
\multirow{2}{*}{$\sqrt{3} \times \sqrt{3}$ VBC} & iPEPS                           & \multirow{2}{*}{} & \multirow{2}{*}{-0.4111} & \multirow{2}{*}{$[0.35,0.42]$}          \\ \cline{2-2}
                                                & 9-PESS                          &                   &                          &                                         \\ \hline
\multirow{2}{*}{$\sqrt{3} \times \sqrt{3}$ VBC} & \multirow{2}{*}{VMC}            & $L=6$             & \multirow{2}{*}{-0.4184} & $[0.31,0.47]$                           \\ \cline{3-3} \cline{5-5} 
                                                &                                 & $L=12$            &                          & $[0.31,0.47]$                           \\ \hline
$3 \times 3$ VBC A                              & \multirow{2}{*}{NQS}            & $L=6$             & -0.42069                 & $[0.23,0.57]$                           \\ \cline{1-1} \cline{3-5} 
$3 \times 3$ VBC B                              &                                 & $L=9$             & -0.42122                 & $[0.24,0.56]$                           \\ \hline
\textbf{27-site SDW}                              & \multirow{2}{*}{iPEPS} & $D=15$            & \textbf{-0.42204(1)}     & \multirow{2}{*}{\textbf{$[0.25,0.46]$}} \\ \cline{3-4}
(This work)                                     &                                 & $D\to\infty$      & \textbf{-0.4238(2)}      &                                         \\ \hline
\end{tabular}
\caption{The values of $e(m)$ for candidate $1/9$ plateau states, as reported in previous studies and in this work, are summarized. The corresponding references are as follows: $Z_3$ QSL~\cite{He2024}; $\sqrt{3}\times\sqrt{3}$ VBC from iPEPS and $9$-PESS~\cite{Fang2023}; $3\times 3$ VBC from~\cite{Cheng2026}; and $3\times3$ VBCs from NQS~\cite{Raikos2026}. The results for this work are obtained using the $27$-site iPEPS ansatz.}\label{tab:comparison}
\end{table}

\begin{figure*}[t]
    \includegraphics[width=0.9\textwidth]{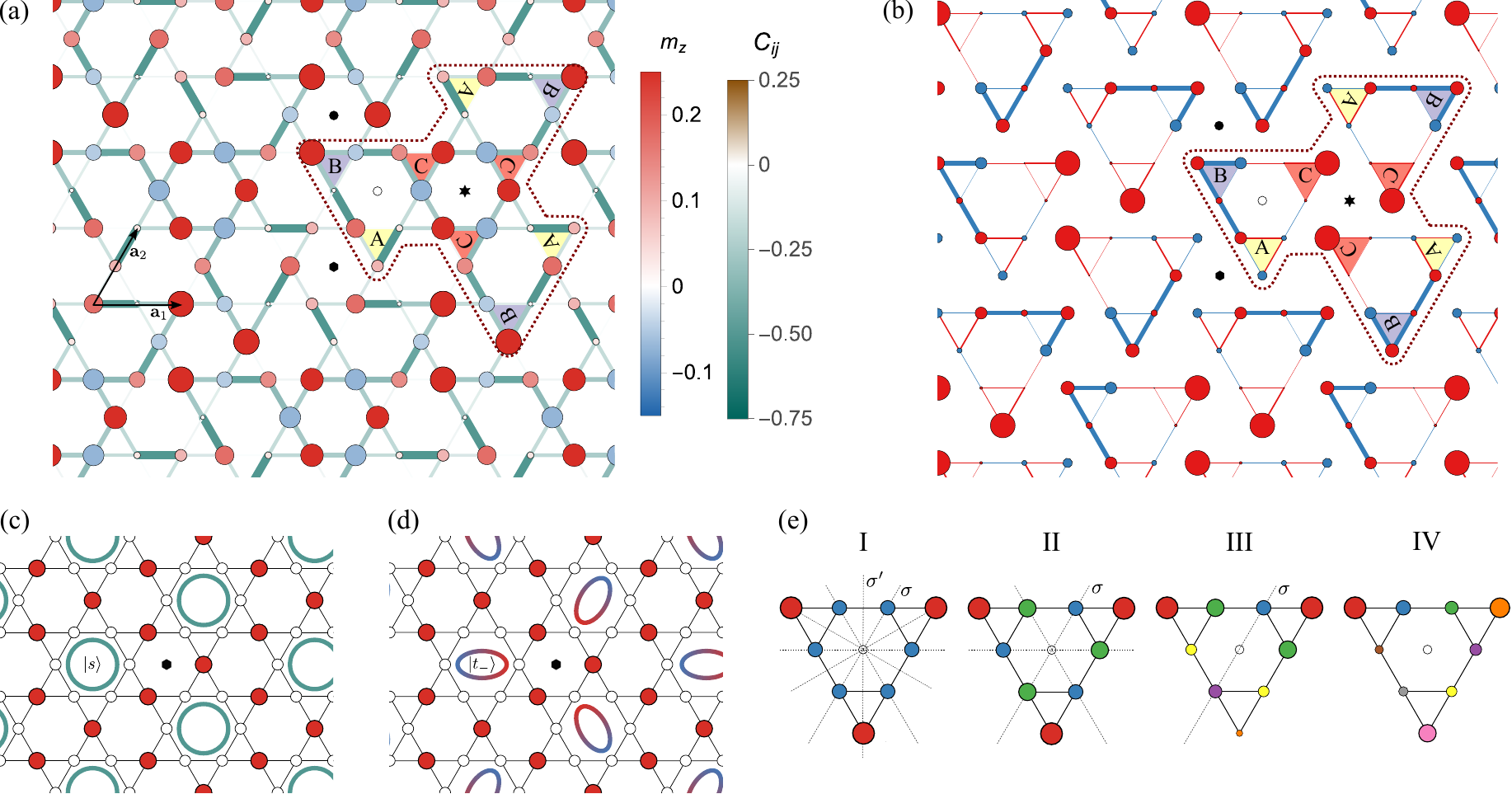}
\caption{Structure of the $1/9$ plateau state as determined by (a) the $27$-site unit-cell tensor network ansatz with bond dimension $D=15$ at a magnetic field $h=0.4$, and (b) the $9$-site CMFT analysis.
The size and color of each circle represent the local magnetization $m_z$, while the thickness and color of each bond correspond to the spin correlation $C_{ij}$. For panel (a), the relevant color scales are provided. In panel (b), blue and red indicate negative and positive values, respectively. The 27-site magnetic unit cells are delineated by dashed lines. Both states exhibit three inequivalent centers of threefold rotation, indicated by filled symbols. The complete pattern is generated by a $9$-site triangular region and its successive rotations by $2\pi/3$ about the respective centers. Panels (c) and (d) show schematic representations of the $m/m_s=1/3$ and $m/m_s=1/9$ plateau states, respectively. The $1/3$ plateau can be approximated by singlets marked by green circles localized on the hexagons, with the remaining spins fully polarized. In the $1/9$ plateau, these singlets are replaced by triplets $|t_-\rangle$. The presence of triplets breaks the sixfold rotational symmetry of the central hexagon, resulting in a nonuniform distribution of local magnetizations. Panel (e) presents a schematic depiction of possible symmetry properties of the supertriangle. The colors and sizes of the circles are selected arbitrarily to distinguish different symmetry scenarios. (I) The $C_6$-symmetric case, characterized by two distinct local magnetizations, as observed for the $7/9$, $5/9$, and $1/3$ plateaus. (II) A scenario in which all $\sigma$ reflections are preserved, while $\sigma'$ reflections are broken. (III) A scenario where two of the $\sigma$ reflection symmetries are further broken, resulting in six independent local magnetizations. (IV) A scenario in which all reflection and sixfold rotational symmetries are absent, allowing for nine independent local magnetizations.}
 \label{fig-structure}
\end{figure*}

Figure~\ref{fig-structure}(a) shows a representative example of local magnetizations and NN spin-spin correlations corresponding to the $1/9$ plateau for $h=0.4$ at $D=15$.
This state exhibits an SDW pattern with a $27$-site magnetic unit cell of the $3\times3$ type, which remains robust for $D \geq 10$. 
Our initial simulations use six independent tensors, allowing all 27 sites in the unit cell to evolve independently. However, since we observe an approximate threefold rotational symmetry about the center of the tensor network unit cell, we exploit this symmetry to reduce the number of independent sites from 27 to 9, thereby improving numerical convergence and obtaining a lower-energy state. The resulting state, thus, can be constructed from a $9$-site triangular cluster by applying successive $\mp2\pi/3$ rotations about the center, combined with translations along the $\mathbf{a}_1\pm\mathbf{a}_2$ directions, where $\mathbf{a}_1$ and $\mathbf{a}_2$ are the primitive lattice vectors. The state contains three inequivalent $C_3$ rotation centers, indicated by filled markers in Figure~\ref{fig-structure}(a).
Although it shares certain similarities with the previously reported VBC states with a $27$-site unit cell~\cite{Cheng2026,Raikos2026}, our state exhibits different local magnetization patterns. Note that the states identified as VBCs in previous studies may also be interpreted as SDWs, given the pronounced spatial modulation of their local magnetizations. Like our state, the VBC reported in Ref.~\cite{Cheng2026} and the two states referred to as VBC A and VBC B in Ref.~\cite{Raikos2026} all feature three inequivalent $C_3$ rotation centers, around which the spins exhibit alternating degrees of polarization. However, on the hexagon marked by the filled hexagon, all spins in our state are polarized along the applied field, whereas in the VBCs reported in Refs.~\cite{Cheng2026,Raikos2026}, the weakly polarized spins point opposite to the field. The distinction from VBC B in Ref.~\cite{Raikos2026} also lies in the arrangement of strongly and weakly polarized spins around the three hexagons that contain the $C_3$-symmetric centers. In VBC B, these arrangements are equivalent around all three hexagons. In our state, however, the arrangement around the hexagon marked by the filled hexagon is reversed relative to those around the hexagons marked by the filled star and filled circle.


Although the $1/9$ plateau exhibits a distinct structure as compared to higher plateaus, it can be approached from the localized-magnon picture. Starting from saturation, the $7/9$, $5/9$, and $1/3$ plateaus correspond to localized one-, two-, and three-magnons on nonoverlapping hexagons [Fig.~\ref{fig-structure}(c)]. The localization of these magnons maintains the sixfold rotational symmetry of each hexagon. Specifically, the $1/3$ plateau may be interpreted as singlets (three-magnon states) on nonoverlapping hexagons, with all other spins remaining polarized [Fig.~\ref{fig-structure}(c)]. One possible route to the $1/9$ plateau is to demote these singlets to $|t_-\rangle$ triplets with $m_z=-1$. The central question is whether introducing these triplets leads to spontaneous breaking of the sixfold symmetry around the hexagons.

To evaluate this scenario, we perform CMFT analysis on clusters of three adjacent down-pointing elementary triangles, hereafter referred to as ``supertriangles'' [Fig.~\ref{fig-structure}(b)]. Motivated by the iPEPS results, the lattice is tiled with copies of the cluster, each related by successive $\mp2\pi/3$ rotations about hexagon centers and translations along $\mathbf{a}_1\pm\mathbf{a}_2$.
In this approach, each supertriangle is solved exactly within a fixed total-$S^z$ sector, while intercluster couplings are treated through self-consistent boundary mean fields.
Although the cluster Hamiltonian respects $C_3$ and $\sigma’$ reflection symmetries, the CMFT solution is not constrained to preserve them and may exhibit additional symmetries in the full lattice state [Fig.~\ref{fig-structure}(e)].

For the $7/9$, $5/9$, and $1/3$ plateaus, the mean-field ground states retain $C_6$ symmetry about the supertriangle centers and display only two distinct local magnetizations~\cite{SupplementalMaterial}, consistent with localized-magnon states [Fig.~\ref{fig-structure}(e)I]. In contrast, at $m/m_s=1/9$, the ground state spontaneously breaks the $C_6$ symmetry [Fig.~\ref{fig-structure}(e)IV] and results in nine distinct local magnetizations within the cluster [Fig.~\ref{fig-structure}(b)].
Consequently, the state also breaks all reflection symmetries of the supertriangle [Fig.~\ref{fig-structure}(e)IV].
The $C_3$ symmetry about the centers indicated by filled symbols in Fig.~\ref{fig-structure}(b) is preserved by construction. Although quantitative agreement with iPEPS is not expected at this level of approximation, the CMFT pattern replicates its key qualitative features: nine distinct local magnetizations and an identical relative arrangement of strongly polarized spins around the hexagons containing the $C_3$ centers.  

This qualitative agreement indicates that the $|t_-\rangle$ states on the supertriangle hexagons reduce the system's energy through spontaneous breaking of $C_6$ symmetry, which redistributes spin polarization into the pattern depicted in Fig.~\ref{fig-structure}(a).
We emphasize that the iPEPS results are not fully converged with respect to $D$, even at $D=15$ [see Fig.~\ref{fig-energy}(c)], so it remains possible that additional symmetries could be restored in the limit $D\to\infty$. Specifically, the weak breaking of the $\sigma$ reflections passing through the hexagonal plaquette marked with a filled hexagon in Fig.~\ref{fig-structure}(a) may vanish at larger $D$, resulting in a state composed of supertriangles with the same symmetries as in Fig.~\ref{fig-structure}(e)III. This scenario would reduce the number of independent local magnetizations from nine to six~\footnote{In Ref.~\cite{Raikos2026}, additional symmetrization was imposed to obtain five independent magnetizations. Neither our numerical results nor our symmetry analysis provides evidence supporting this additional symmetrization.}.  Our finite-$D$ results cannot confirm or exclude this possibility [see Supplemental Material~\cite{SupplementalMaterial}]. Nevertheless, we find no indication of a restoration of all $\sigma$ reflections while the $\sigma'$ reflections remain broken, as in Fig.~\ref{fig-structure}(e)II. Additionally, CMFT calculations using a $27$-site cluster~\cite{SupplementalMaterial} exhibit the same symmetry breaking and yield nine distinct local magnetizations. These findings suggest that the $1/9$ plateau most likely exhibits nine distinct local magnetizations, although the possibility of six cannot be entirely ruled out. 

In summary, the combination of iPEPS and CMFT leads to a consistent picture of the $1/9$ magnetization plateau of the spin-$1/2$ KHA as a SDW that breaks sixfold rotational symmetry and consists of three $9$-site clusters related by $\pm2\pi/3$ rotations, resulting in a 27-site magnetic unit cell with nine distinct local magnetizations.
An ideal technique to test this prediction is NMR, which provides a histogram of the local magnetizations, or some linear combinations if the nuclear spin is not on site, as in the $^{35}$Cl NMR investigation of \ce{Y}-kapellasite~\cite{chatterjee2026gapped19magnetizationplateau}. Since the $1/9$ plateau has been observed at fields of approximately $20$ $\mathrm{T}$ in yttrium-based hydroxyhalides, \ce{Y}-kapellasite, and titanium-based kagome compounds at fields ~\cite{Jeon2024,Suetsugu2024,q3yj-t2hj,Zheng2025,hayashi202659magnetizationplateauspinsupersolidity,chatterjee2026gapped19magnetizationplateau,chaudhary2026universality19magnetizationplateau}, this prediction should be accessible to experimental verification. 
However, further theoretical studies of the spin-1/2 KHA with modified interactions are needed to clarify the microscopic origin of the robust $1/9$ plateau-like phases and investigate the possible role of bond anisotropy~\cite{Jeon2024,Suetsugu2024,q3yj-t2hj,Zheng2025,hayashi202659magnetizationplateauspinsupersolidity,chatterjee2026gapped19magnetizationplateau,chaudhary2026universality19magnetizationplateau}. A comparable analysis of the spin-$1$ KHA and related systems would also be interesting. In these systems, a gapped trimerized ground state has been identified~\cite{Changlani-Spin-1_Kagome,Ghosh2016,Ghosh2018}, whereas no $1/9$ plateau was observed in tensor network calculations with a $9$-site unit cell~\cite{Picot2016}.  Work is in progress along these lines. 

At the technical level, the critical next step involves applying advanced tensor network optimization techniques, including full-update and automatic-differentiation methods, to the $27$-site ansatz, and more generally to enlarged-unit-cell ansatzes, to test ground-state structures with improved accuracy. It is therefore particularly valuable to develop strategies that enable these calculations at large bond dimensions. More broadly, our results demonstrate that tensor networks with enlarged unit cells, combined with hexagonal CTMRG, are effective for investigating phases with complex spatial structures. This methodology could also be extended to the maple-leaf lattice in a magnetic field~\cite{Ghosh2023,Ghosh2024-al}, where enlarged unit cells may facilitate access to exotic magnetization plateaus and field-induced states with spatial structures not captured in smaller-unit-cell calculations.

\textit{Acknowledgments:}
We thank Chisa Hotta and Sylvain Capponi for useful discussions. 
The work is supported by the Swiss National Science Foundation Grant No. 212082.
Numerical computations were performed using the facilities of the Scientific IT and Application Support Center of EPFL (SCITAS).

The tensor network calculations in this work were performed using the ITensor library~\cite{itensor}.

The data supporting the findings of this letter are openly available \cite{Ghosh2025_data}.

\bibliography{Refs.bib}

\end{document}


\title{Supplemental Material for
``Theory of the $1/9$ Magnetization Plateau of the Spin-$1/2$ Kagome Heisenberg Antiferromagnet''}

\author{Pratyay Ghosh}
\email{pratyay.ghosh@epfl.ch}
\affiliation{Institute of Physics, Ecole Polytechnique Fédérale de Lausanne (EPFL), CH-1015 Lausanne, Switzerland}

\author{Fr\'ed\'eric Mila}
\affiliation{Institute of Physics, Ecole Polytechnique Fédérale de Lausanne (EPFL), CH-1015 Lausanne, Switzerland}

\maketitle

\section{Infinite projected entangled pair states (iPEPS)}
\label{sec:methods}

We use infinite projected entangled pair states (iPEPS) to approximate the ground states of the infinite kagome lattice. The iPEPS ansatz represents the many-body wave function as an infinite two-dimensional tensor network, with each tensor carrying one physical index and three virtual indices. The physical index has dimension $d=2$ for spin-$1/2$ degrees of freedom, while the virtual indices have dimension $D$, which controls the amount of entanglement that can be represented. All tensor-network calculations in this work were performed using the ITensor library~\cite{itensor}.

To investigate the spatial structure of the magnetization plateau, we employ enlarged unit cells containing 27 kagome sites. The 27-site calculations are performed up to $D=15$. The enlarged unit cells allow the tensors to develop spatially nonuniform patterns of local magnetization and nearest-neighbor correlations without imposing the translational symmetries of smaller unit cells.

The tensors are optimized using imaginary-time evolution with the simple-update (SU) algorithm~\cite{Jiang2008}. Starting from randomly initialized tensors, we apply a sequence of local imaginary-time evolution operators,
\begin{equation}
    |\psi_{\mathrm{GS}}\rangle
=
\lim_{\beta\rightarrow\infty}
e^{-\beta H}|\psi_0\rangle
\nonumber\
\simeq
\lim_{n\rightarrow\infty}
\left[
\prod_{\langle i,j\rangle}
e^{-\tau H_{ij}}
\right]^n
|\psi_0\rangle ,
\end{equation}

where $\tau$ is the imaginary-time step and $H_{ij}$ denotes the local Hamiltonian acting on a nearest-neighbor bond. In practice, we gradually reduce $\tau$ during the evolution. After applying each two-site gate $e^{-\tau H_{ij}}$, the resulting tensors are truncated back to bond dimension $D$ using a singular-value decomposition. The SU procedure provides an efficient optimization of the tensors and is particularly useful for exploring the large parameter space associated with the enlarged unit cells.

After optimization, physical observables are evaluated by contracting the corresponding infinite tensor network. We use the corner transfer matrix renormalization group (CTMRG) method to construct an effective environment around the tensors~\cite{Nishino1996,Orus2009}. For the kagome lattice, the contraction is formulated directly in terms of the lattice geometry, avoiding the introduction of auxiliary lattice deformations. The environment is characterized by a boundary bond dimension $\chi$, which controls the accuracy of the contraction. Observables are obtained by contracting the local tensors with this environment and are extrapolated or checked for convergence with increasing $\chi$.

For each optimized iPEPS, we increase $\chi$ until the relevant observables, in particular the energy and magnetization, are converged within the required numerical accuracy. The CTMRG iterations are repeated after successive imaginary-time evolution steps so that the environment remains consistent with the evolving tensors. For the largest bond dimensions considered here, we use $\chi$ values sufficiently large to ensure that the reported energies and local observables are converged with respect to the environment dimension.

For the plateau state, we analyze the local magnetization and nearest-neighbor spin-spin correlations to characterize its real-space structure.





\section{Further results from cluster mean-field theory}

We also examine the local magnetization patterns obtained from CMFT on the nine-site cluster for the $1/9$, $1/3$, and $5/9$ plateaus [Fig.~\ref{fig:ed-cmft}(a)–(c)]. For the $1/9$ plateau, we also show the corresponding spin-spin correlations in Fig.~\ref{fig:ed-cmft}(d). We compare these results with the iPEPS calculations in the main text.

\begin{figure}[ht]
    \centering
    \includegraphics[width=\columnwidth]{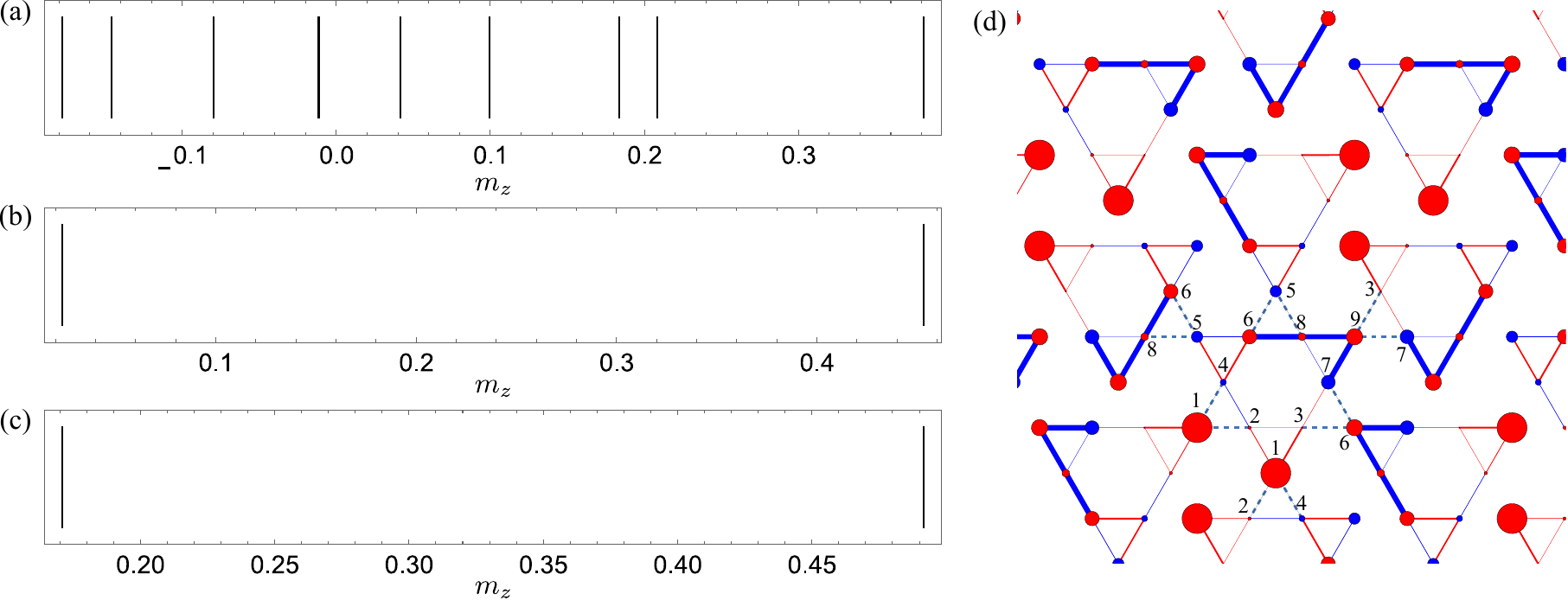}
    \caption{%
    (a)–(c) Local magnetizations on the nine-site cluster obtained from CMFT for the (a) $1/9$, (b) $1/3$, and (c) $5/9$ magnetization plateaus. (d) Local magnetizations and spin-spin correlations for the $1/9$ magnetization plateau. Blue (red) indicate negative (positive) values. We show the nine-site cluster and the interactions treated with mean-field are shown by dashed lines.
}
    \label{fig:ed-cmft}
\end{figure}

To further investigate the structure of the $1/9$ magnetization plateau, we study a 27-site kagome cluster using exact diagonalization (ED). Figure~\ref{fig:ed-cmft}(a) shows the lowest energies in the $m_z=1/2$, $3/2$, and $5/2$ sectors as a function of magnetic field $h$. Over a finite field range, the $m_z=3/2$ sector has the lowest energy, corresponding to the $1/9$ plateau.

We examine the real-space structure of this state through the local magnetization and nearest-neighbor spin-spin correlations, $C_{ij}=\langle\mathbf S_i\cdot\mathbf S_j\rangle$. The resulting correlation pattern is shown in Fig.~\ref{fig:ed-cmft}(b), using the same bond-color and bond-width convention as in the main text. The ED result displays a pronounced spatial modulation of the nearest-neighbor correlations.

As an independent check, we also perform cluster mean-field theory (CMFT) using the same 27-site cluster. In CMFT, the intra-cluster interactions are treated explicitly, while the coupling to neighboring clusters is decoupled at the mean-field level and determined self-consistently. The resulting nearest-neighbor correlations are shown in Fig.~\ref{fig:ed-cmft}(c). The CMFT calculation reproduces the characteristic spatial modulation found in ED, demonstrating that the observed correlation pattern is robust against the treatment of the inter-cluster couplings. Both approaches therefore yield a consistent real-space structure for the $1/9$ plateau.

\begin{figure}[ht]
    \centering
    \includegraphics[width=\columnwidth]{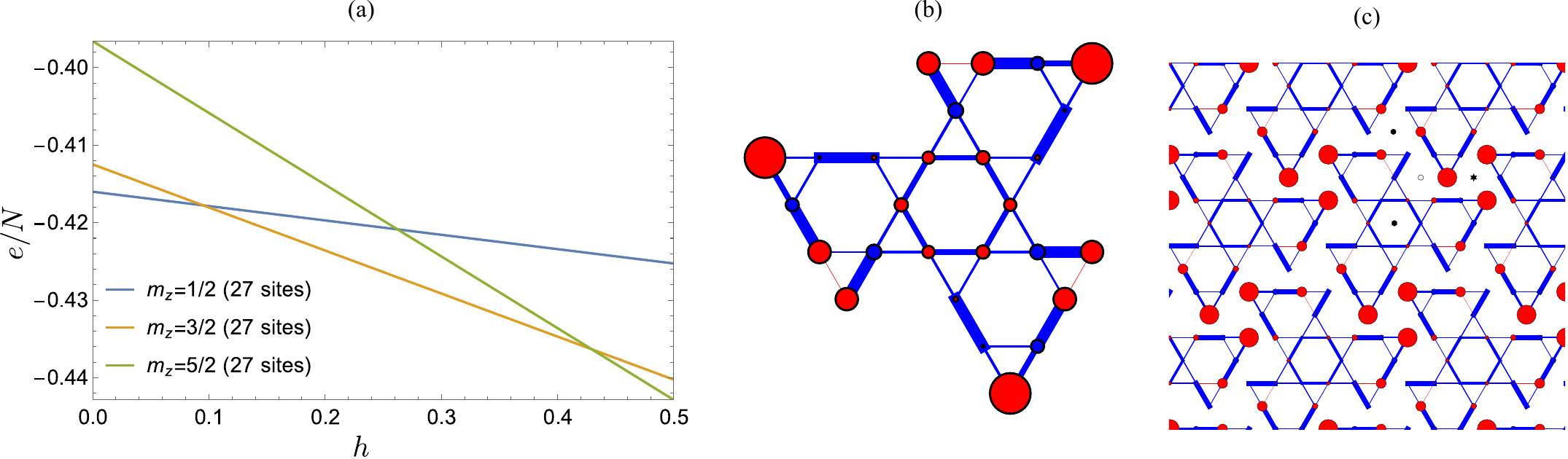}
    \caption{%
    Exact-diagonalization (ED) and cluster mean-field theory (CMFT) results for the 27-site kagome cluster. (a) ED energies in the $m_z=1/2$, $3/2$, and $5/2$ sectors as a function of magnetic field $h$, showing a finite field range in which the $m_z=3/2$ sector has the lowest energy. (b) Real-space pattern of the local magnetization and nearest-neighbor spin-spin correlations, $C_{ij}=\langle\mathbf S_i\cdot\mathbf S_j\rangle$, obtained from ED in the $m_z=3/2$ sector. (c) Corresponding correlations obtained from CMFT. Blue (red) indicate negative (positive) values. The ED and CMFT results show the same characteristic spatial pattern of correlations associated with the $1/9$ magnetization plateau.
}
    \label{fig:ed-cmft}
\end{figure}

\section{Further iPEPS results}
\begin{figure}[ht]
    \centering
    \includegraphics[width=0.6\columnwidth]{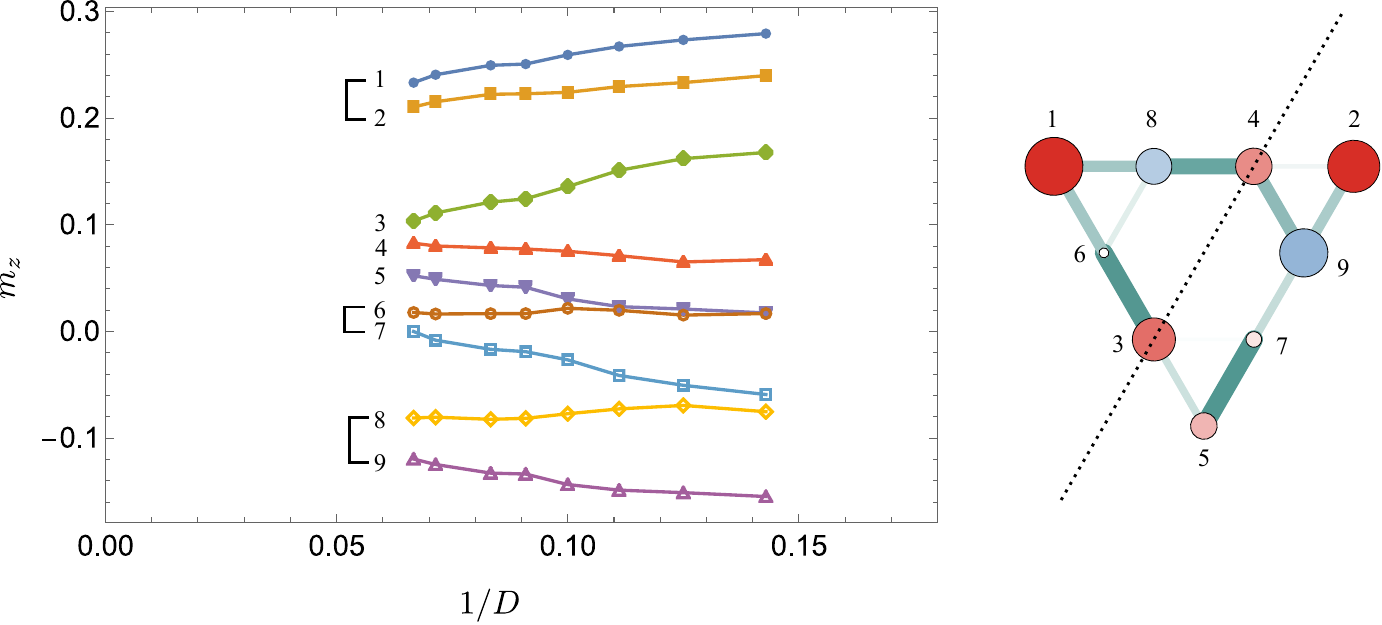}
    \caption{%
    Evolution of the local magnetizations of the $1/9$ plateau state as a function of $1/D$. Although a reliable extrapolation to the $D\to\infty$ limit is not possible, the results show no clear indication of a restoration of the sixfold symmetry as $D$ increases. However, the weak breaking of the reflection symmetries along the dashed line leaves open the possibility that some symmetries may be restored at larger $D$. The local magnetizations that may become equal in the \(D\to\infty\) limit, leaving only six independent values, are also indicated.
}
    \label{fig:ipeps}
\end{figure}

To further characterize the $1/9$ plateau, we examine the evolution of the local magnetizations with increasing bond dimension $D$. As shown in Fig.~\ref{fig:ipeps}, the local magnetizations exhibit no clear tendency toward the restoration of sixfold rotational symmetry. Although a reliable extrapolation to the $D\to\infty$ limit is not possible, the observed behavior provides no clear indication that the symmetry breaking vanishes in this limit.

\begin{figure}[ht]
    \centering
    \includegraphics[width=\columnwidth]{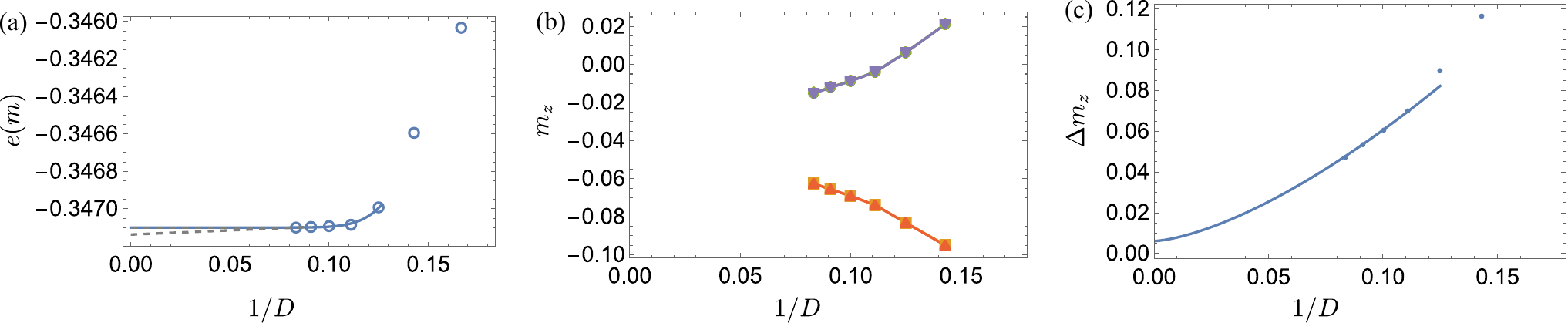}
    \caption{%
    (a) Energy as a function of $1/D$ for the $1/3$. (b) Local magnetizations on the hexagons hosting localized magnons for the $1/3$ plateau. The observed symmetry breaking arises from the construction of the tensor-network ansatz. (c) Difference between the local magnetizations, which decreases with increasing $D$. This indicates the restoration of sixfold symmetry of the plateau in the $D\to\infty$ limit.
    }
    \label{fig:ipeps_2}
\end{figure}

For comparison, we perform the same analysis for the $1/3$ and $5/9$ plateaus, which can be understood in terms of localized magnons on nonoverlapping hexagonal plaquettes. In these cases, the tensor-network ansatz introduces a small symmetry breaking in the local magnetizations on the hexagons hosting the localized magnons. As shown in Figs.~\ref{fig:ipeps} and \ref{fig:ipeps_2}, the differences between the local magnetizations decrease with increasing $D$, indicating that the sixfold rotational symmetry is progressively restored in the $D\to\infty$ limit. This behavior contrasts with that of the $1/9$ plateau, where the local magnetizations show no analogous tendency toward symmetry restoration.

\bibliography{Refs.bib}